\documentclass[11pt,a4paper,final]{article}
\usepackage[utf8]{inputenc}
\usepackage{authblk}
\usepackage{setspace}
\usepackage[margin=1.25in]{geometry}
\usepackage{graphicx}
\usepackage{subcaption}
\usepackage{amsmath}
\usepackage{amsfonts}
\usepackage{amssymb}
\usepackage{float}
\usepackage{etoolbox}
\usepackage{adjustbox}
\usepackage{enumitem}
\usepackage[update,prepend]{epstopdf}
\usepackage{color,colortbl,soul}
\usepackage{rotating}
\usepackage{adjustbox}
\usepackage{xspace} 
\usepackage{upgreek}
\usepackage{array}
\usepackage[numbers,sort&compress]{natbib}
\makeatletter
\edef\orig@output{\the\output}
\makeatother% Fixing an inflexible document class by adding a finite (1pt) stretch on every page
\usepackage{hyperref}
\hypersetup{
    breaklinks=true,
    colorlinks=true,
    linkcolor=blue,
    citecolor=blue,      
    urlcolor=blue,
}
\usepackage{xurl}
\newcolumntype{s}{>{\columncolor[HTML]{AAACED}} p{3cm}}
\arrayrulecolor{black}
\usepackage[T1]{fontenc}
\usepackage{footnote}
\usepackage{bm}% bold math
\usepackage{cancel}
\usepackage{mathrsfs}
\usepackage{etoolbox}
\apptocmd{\sloppy}{\hbadness 10000\relax}{}{}
\usepackage{pifont}% For special characters
\usepackage{lineno}
\usepackage{xcolor}
\usepackage{tcolorbox}

\tcbset{
  colback=blue!5!white,      % <-- Change shade color here
  colframe=blue!5!white,    % <-- Change border color here
  boxrule=0mm,             % Completely removes the border line thickness
  width=0.95\textwidth,      % <-- Change box width here
  center,                    % Centers the box because it's narrower
  arc=0mm,                   % Corner roundness
  left=0.0mm, right=0.0mm,       % Inside spacing side-to-side
  top=1mm, bottom=1mm        % Inside spacing top-and-bottom
}
\usepackage{tikz,xcolor}
\definecolor{lime}{HTML}{A6CE39}
\DeclareRobustCommand{\orcidicon}{%
	\begin{tikzpicture}
	\draw[lime, fill=lime] (0,0) 
	circle [radius=0.16] 
	node[white] {{\fontfamily{qag}\selectfont \tiny ID}};
	\draw[white, fill=white] (-0.0625,0.095) 
	circle [radius=0.007];
	\end{tikzpicture}
	\hspace{-2mm}
}

\foreach \x in {A, ..., Z}{%
	\expandafter\xdef\csname orcid\x\endcsname{\noexpand\href{https://orcid.org/\csname orcidauthor\x\endcsname}{\noexpand\orcidicon}}
}

\definecolor{BrightGray}{rgb}{.94,.94,.94}
\usepackage[english]{babel} % or whatever your language is
\addto\captionsenglish{}

\usepackage{caption}
\title{Harnessing Machine Learning for Ultrafast Science in Facility Settings: Opportunities and Challenges}
\author[1]{Mousumi Upadhyay Kahaly\orcidG{}}
\author[2]{Ishraq Md Anjum\orcidD{}}
\author[3]{Davorin Peceli\orcidC{}}
\author[1]{Zsolt Diveki\orcidB{}}
\author[1]{Subhendu Kahaly\orcidA{}*}
\affil[1]{ELI ALPS, The Extreme Light Infrastructure ERIC, Wolfgang Sandner u. 3., 6728 Szeged, Hungary}
\affil[2]{Department of Computer Science and Electrical Engineering, University of Maryland Baltimore County, Baltimore, MD 21250, USA}
\affil[3]{ELI Beamlines Facility, The Extreme Light Infrastructure ERIC, Za Radnici 835, 252 41 Doln\'{\i} B\v{r}e\v{z}any, Czech Republic}
\affil[*]{Email: \textcolor{blue}{subhendu.kahaly@eli-alps.hu}}

\date{} % This leaves the date empty so it won't display
\begin{document}
\sloppy
\maketitle
% \date{\today} 

%\linenumbers

%%%%%% Abstract %%%%%%%%%%%%%%%%%%%%%%%%%%%%%%%%%
\begin{tcolorbox}
\begin{abstract}
Ultrafast science is rapidly evolving from a collection of specialized laser experiments into a facility-scale, data-intensive research endeavour. High-repetition-rate femtosecond and attosecond sources, synchronized multidimensional diagnostics, and increasingly nonlinear laser-matter interactions now generate heterogeneous data streams whose volume, speed, and complexity often exceed the capacity of conventional manual workflows. This chapter examines how machine learning can become an enabling layer for next-generation ultrafast facilities, with emphasis on the Extreme Light Infrastructure (ELI) ecosystem. We discuss the scientific drivers for artificial-intelligence-assisted experimentation, propose a facility-scale digital architecture spanning acquisition, ingestion, storage, analytics, machine learning, adaptive feedback, and user interfaces, and review representative examples from ELI ALPS, ELI Beamlines, and ELI-NP. The central thesis is that machine learning should not be treated merely as an offline data-analysis tool; rather, when constrained by physical insight and embedded in robust data infrastructure, it can support real-time diagnostics, surrogate modeling, anomaly detection, experimental optimization, and progressively autonomous discovery workflows.

\end{abstract}
\end{tcolorbox}
%%%%%%%%%%%%%%%%%%%%%%%%%%%%%%%%%%%%%%%%%%%%%%%%%%

%%%%%% Main Text %%%%%%%%%%%%%%%%%%%%%%%%%%%

\section{Introduction: Why Ultrafast Facilities Need Machine Learning}
Over the last two decades, advances in high-power and high-repetition-rate laser technology \cite{Gizzi2026,Hu2026,Veisz2025} have transformed ultrafast science from a predominantly laboratory-scale discipline into a data-intensive facility enterprise with a broad range of scientific and societal applications \cite{Phillips2015,Borsch2023,Hideghty2025}. Contemporary extreme-light infrastructures routinely access femtosecond and attosecond time scales, relativistic laser–plasma regimes, high-field secondary-source generation, and nonlinear response in gases, plasmas, molecules, nanostructures, and condensed phases \cite{Charalambidis2017,Kahaly2025,Tomimasu1998}. At the same time, experimental campaigns increasingly combine detector images, spectra, particle diagnostics, timing signals, environmental metadata, and machine-state information acquired at high repetition rate. The result is not only a growth in data volume but a qualitative change in scientific complexity: experiments are now characterized by high-dimensional control spaces, nonlinear correlations, shot-to-shot fluctuations, and competing sources of noise and drift.

Traditional workflows based on manual optimization, delayed offline analysis, and expert intuition remain indispensable, but they are increasingly insufficient as the dominant operating paradigm. The scientific bottleneck is shifting from data acquisition to data interpretation, decision-making, and adaptive control. Consequently, the emergence of machine learning (ML), artificial intelligence (AI) and autonomous experimentation capabilities has aligned with a transformative phase in ultrafast science, where data-driven approaches are transitioning from optional tools to essential components of facility operations. AI-ML provides a route to close this gap by extracting latent structure from high-dimensional data, building fast surrogate models for expensive simulations, recognizing anomalous machine states, and guiding experimental parameters toward physically informative regions. 

The need for AI-ML in ultrafast science arises from several mutually reinforcing trends. First, high-order harmonic generation \cite{Agostini2004,Nayak2019}, attosecond spectroscopy, laser-wakefield acceleration \cite{Esarey2009}, laser-driven ion acceleration \cite{Macchi2013}, and ultrafast condensed-matter investigations \cite{delaTorre2021,Zong2023} involve strongly coupled variables such as pulse energy, duration, wavefront, carrier-envelope phase (CEP), target density, gas pressure, delay, polarization, and phase-matching conditions \cite{Weissenbilder2022,Appi2023}. Second, high-repetition-rate target and detector technologies \cite{Prencipe2017,BalazsN2023} create data streams that must be analyzed with minimal latency if beam time is to be used efficiently. Third, ultrafast facilities increasingly serve multidisciplinary communities whose experiments combine strong-field physics, chemistry, plasma science, materials science, and photonics. The latter trend emphasizes the need for an integrated, multiscale and multiphysics approach to data acquisition and analysis. This also highlights the need of employing Physics informed machine learning (PIML) \cite{Karniadakis2021,Meng2025} that embeds domain knowledge appropriately, in a facility setting.

For ultrafast science, the most valuable role of AI-ML is therefore not only generic automation, but more importantly physically constrained acceleration of discovery. The eventual long term aim would be to reach a state where end to end AI scientists \cite{Lu2026} would work in collaboration with human scientists under explicit risk-mitigation and oversight mechanisms \cite{Tang2025}. Such applications can only be appropriately done within the data rich environment provided in an ultrafast laser facility setting. The Extreme Light Infrastructure illustrates this transition particularly clearly.
%%%%%%%%%%%%%%%%%%%%%%%%%%%%%%%%%%%%%%%%%%%%%%%%%%%%%%%%%%%%%%%%%%%%%%%%%%%%%%%%%%%%%%%%%%%%%%%%%%%%%%%%%%%%%%%%%%
\begin{figure*}[t]
    \centering
    \includegraphics[width=1\linewidth]{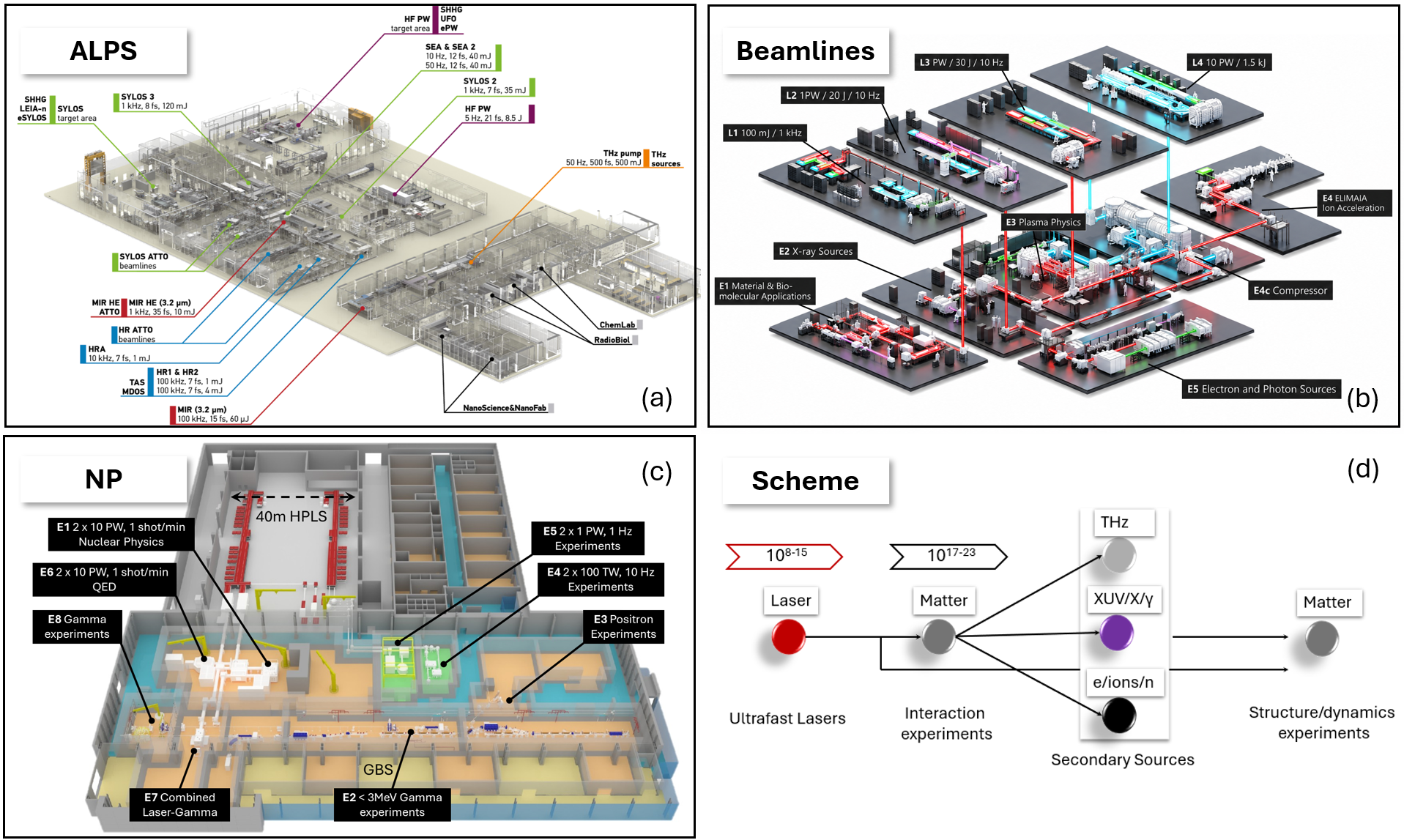}
    \caption{\textbf{Overview of the ELI ecosystem.} Panels (a)–(c) summarize the scientific focus and principal laser capabilities of ELI ALPS (Szeged, Hungary), ELI Beamlines (Prague, Czech Republic), and ELI-NP (Măgurele, Romania), respectively. Panel (d) presents a generalized experimental workflow shared across the three facilities. Ultrashort laser pulses with different durations, repetition rates, pulse energies, central wavelengths, and peak intensities (in $Wcm^{-2}$ indicated inside the arrows) interact with gases, plasmas, liquids, solids, or nanostructured materials. The interaction may be interrogated directly to resolve ultrafast dynamics, or the resulting secondary photon and particle sources may be used in subsequent experiments.}
    \label{fig1}
\end{figure*}
%%%%%%%%%%%%%%%%%%%%%%%%%%%%%%%%%%%%%%%%%%%%%%%%%%%%%%%%%%%%%%%%%%%%%%%%%%%%%%%%%%%%%%%%%%%%%%%%%%%%%%%%%%%%%%%%%%%

\section{Extreme Light Infrastructures: Scientific opportunities for data-intensive ultrafast science}

The Extreme Light Infrastructure \cite{ELIERIC} represents one of the most significant international investments in the direction of a user facility supporting ultrafast laser matter interaction. Established as a pan-European research infrastructure, ELI was conceived to provide researchers with access to laser systems operating at the frontier of intensity, pulse duration, and repetition rate. Rather than concentrating all capabilities within a single laboratory, ELI adopted a distributed architecture consisting of three complementary pillars: ELI ALPS \cite{ELIALPS} (Attosecond Light Pulse Source) in Szeged, Hungary; ELI Beamlines \cite{ELIBL} in Doln\'{\i} B\v{r}e\v{z}any near Prague, Czech Republic; and ELI Nuclear Physics \cite{ELINP} (ELI-NP) in M\u{a}gurele, Romania. Although united by a common scientific vision centered on extreme light-matter interactions, each pillar addresses a distinct segment of the experimental landscape. Together they form a uniquely diverse ecosystem that spans attosecond science, high-field laser-plasma physics, secondary particle and radiation source development, and nuclear photonics.

The differences among the three facilities are reflected not only in their scientific missions but also in the operational characteristics of their experiments. ELI ALPS \cite{Charalambidis2017,Shirozhan2024,Kahaly2025} is optimized for high-repetition-rate ultrafast measurements and attosecond spectroscopy, where statistical precision, timing stability, and large experimental datasets are central requirements. ELI Beamlines \cite{Rus2011,Cimmino2023,Gizzi2026} focuses on high-power laser systems and laser-driven secondary sources, including X-rays, electrons, protons, and other particle beams, serving a broad multidisciplinary user community. ELI-NP \cite{Ur2025} combines multi-petawatt laser technology with a high-brilliance gamma-beam system to explore photonuclear interactions and strong-field nuclear physics. Despite these differences, the facilities share common technological foundations, including advanced laser architectures, synchronized diagnostics, distributed control systems, large-scale data acquisition networks, high-performance computing resources, and increasingly complex user experiments. These commonalities make ELI a natural testbed for facility-scale machine learning and AI tools that can accelerate analysis, improve facility efficiency, and potentially support autonomous experimental workflows across multiple scientific domains.

Figure~\ref{fig1} provides an overview of the ELI ecosystem. The first three panels Figure~\ref{fig1}(a-c) summarize the major scientific capabilities and experimental focus areas of ELI ALPS, ELI Beamlines, and ELI-NP, respectively. The fourth panel Figure~\ref{fig1}(d) presents a simplified scheme describing how experiments are organized in all three facilities. Lasers with different ultrashort pulse durations, repetition rates, pulse energies, and central wavelengths enable interactions across distinct phases of matter, accessing a wide range of peak‑intensity regimes. This makes it possible either to probe the ultrafast dynamics of the interaction itself or to harness the resulting secondary sources for further investigation.

Alongside advances in ultrafast light sources, the continuous implementation and integration of sophisticated diagnostics, multidimensional detectors, and high-repetition-rate acquisition systems have dramatically increased the volume, spped and complexity of experimental data. At large-scale facilities such as ELI, these challenges are compounded by high-repetition-rate operation and the simultaneous use of multiple diagnostics, synchronized at the shot level. These require the integration of heterogeneous data streams across instruments, temporal scales, and experimental conditions. Extracting physically meaningful information therefore needs more than scalable data storage. It necessitates a traceable framework for integrating heterogeneous measurements acquired with different dimensionalities, temporal resolutions, noise characteristics, and instrumental response functions.

Beyond increasing data volumes, ultrafast experiments are also becoming scientifically and operationally more complex. Many experiments involve high-dimensional parameter spaces in which several interdependent variables determine the measured response \cite{Weissenbilder2022,Appi2023}. For example, all-optical control of laser–matter interactions and the associated secondary radiation and particle sources requires precise optimization of multiple experimental parameters \cite{GulysOldal2020,GulysOldal2021,DeMarco2023,Shirozhan2026}. Efficient exploration of these coupled control spaces may therefore require AI-ML based optimization, particularly when conventional parameter scans are prohibitively expensive or cannot keep pace with the experimental repetition rate. Similar challenges arise in high-harmonic spectroscopy of quantum systems, where strongly nonlinear interactions can produce subtle spectral signatures \cite{Chatziathanasiou2019,Csizmadia2021,Koll2026}. Machine-learning-assisted approaches could help optimize such experiments and identify these signatures in real time, enabling more effective online interpretation.

Consequently, the rate at which data can be generated and scientific insight extracted
needs to increase rapidly. Contemporary facilities can acquire and store vast quantities of information in real time. However, the processes required to calibrate, analyze, interpret, and ultimately act upon these data remain constrained by computational resources, data-management infrastructures, analysis workflows, and the availability of domain expertise \cite{Dongarra2024}. At the same time, advances in scientific machine learning and data-driven surrogate modeling are opening new pathways for accelerating complex simulations and extracting hidden patterns from highdimensional data. These approaches can thereby enable more efficient scientific discovery \cite{Azizzadenesheli2024}. A related opportunity is the development of numerical or computational models as digital twins of specific experimental implementations. For example, coupled ultrafast thermionic emission and three-dimensional thermal-evolution models \cite{UpadhyayKahaly2024}, developed inhouse, may be employed as a digital twin of laser-induced phase changes near the damage threshold of an optical surface. These possibilities can be tested and implemented in facilities such as ELI, where diverse types of data are available under different experimental operating conditions. 

On the one hand, powerful new AI tools have created unprecedented opportunities to discover and design materials with tailored functional properties \cite{Zeni2025}; on the other hand, emerging autonomous knowledge-representation frameworks \cite{Venugopal2024} have made this vast landscape tractable and systematically navigable. Together, these advances, when coupled with facility-scale ultrafast experimental platforms-open entirely new domains for multi-physics, multi-scale scientific exploration with the potential of adopting autonomous experimental methodologies \cite{Stach2021}.

Ultrafast science increasingly converges with material systems spanning diverse dimensionalities \cite{Krausz2009,Nayak2019}. In nanostructures and complex materials, defect-driven, interface-mediated, and curvature-induced phenomena critically shape emergent electronic and optical responses \cite{Basov2017, Jagadish2024, V2024}. This convergence links the intrinsic complexity of condensed‑matter systems with the probing power of strong‑field and attosecond‑scale techniques \cite{Basov2025}. In parallel, established ultrafast methods are rapidly expanding their reach-uncovering exotic superconducting behavior \cite{Alcal2022}, enabling the exploration of new regimes in strong‑field quantum optics \cite{Bhattacharya2023}, and taking early steps toward ultrafast quantum simulations of entanglement dynamics \cite{Bharti2023}.

Building on these expanding scientific possibilities, ultrafast science facilities provide multiple uniquely enabling platforms for the broader research community. They allow the emergent phenomena in complex and low‑dimensional materials to be interrogated under extreme conditions, thus creating a coherent bridge between condensed‑matter physics, photonics, plasma science, and ultrafast dynamics. In particular, the ELI facilities illustrate how ultrafast science is evolving into a data-intensive environment, where increasingly sophisticated experiments generate unprecedented volumes of information and demand faster pathways to scientific insight. Meeting these challenges requires a unified digital framework that can connect instruments, data streams, computational resources, and intelligent decision-making systems, enabling machine learning and AI to become integral components of future facility operations.

These scientific opportunities share a common operational requirement: an integrated digital architecture capable of connecting synchronized measurements, computational models, and experiment-level decisions. The following section outlines such an architecture for facility-scale ultrafast science.

%%%%%%%%%%%%%%%%%%%%%%%%%%%%%%%%%%%%%%%%%%%%%%%%%%%%%%%%%%%%%%%%%%%%%%%%%%%
\section{A Digital Architecture for Machine-Learning-Enabled Ultrafast Facilities}

\begin{figure*}[t]
    \centering
    \includegraphics[width=1\linewidth]{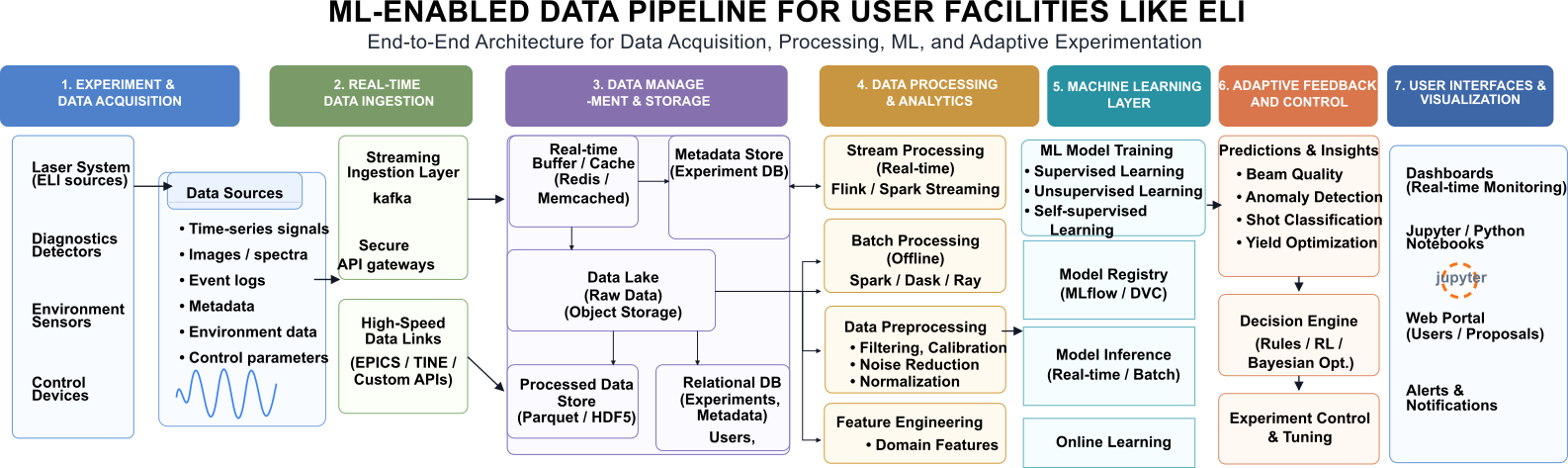}
    \caption{\textbf{Machine-learning-enabled digital ecosystem for facility-scale ultrafast science.} Experimental data generated by laser systems, diagnostics, sensors, and control devices are ingested, stored, processed, and analyzed through a multi-layer architecture. Machine learning models support tasks such as anomaly detection, prediction, optimization, and automated decision-making, while adaptive feedback mechanisms enable intelligent control of experimental conditions. User interfaces provide researchers with real-time access to data products, predictive insights, and facility performance metrics.}
    \label{fig2}
\end{figure*}

The growing complexity of experiments conducted at ELI facilities necessitates a digital infrastructure capable of transforming raw measurements into actionable scientific knowledge in near real time. Figure~\ref{fig2} illustrates a potential machine-learning-enabled data ecosystem designed to support this objective. The architecture is organized into seven interconnected layers that span the complete lifecycle of experimental data, from acquisition at scientific instruments to user-facing visualization and adaptive control. A credible AI-ML strategy for ultrafast science must begin with the data architecture. The proposed facility stack should be organized as a closed-loop online-physical system rather than a linear experiment-to-analysis pipeline. In the following we provide a basic scheme of one of the possible minimal implementations that would allow us to achieve the initial target. The presented scheme illustrates the general fundamental principles.
%%%%%%%%%%%%%%%%%%%%%%%%%%%%%%%%%%%%%%%%%%%%%%%%%%%%%%%%%%%%%%%%%%%%%%%%%%%%%%%%%%%%%%%%%%%%%%%%%%%%%%%%%%%%%
%%%%%%%%%%%%%%%%%%%%%%%%%%%%%%%%%%%%%%%%%%%%%%%%%%%%%%%%%%%%%%%%%%%%%%%%%%%%%%%%%%%%%%%%%%%%%%%%%%%%%%%%%%%%%%%%%%%
The first layer, Experiment and Data Acquisition, represents the physical origin of information within the facility. Laser systems, diagnostic detectors, environmental monitoring instruments, and control devices continuously generate heterogeneous datasets comprising detector images, spectra, time-series measurements, event records, operational metadata, and engineering parameters. In attosecond and high-field experiments, where subtle variations in laser stability or environmental conditions may influence experimental outcomes, preserving these information with high temporal fidelity is essential.

The Real-Time Data Ingestion layer subsequently aggregates these distributed data streams through high-throughput communication infrastructures based on technologies such as Kafka and EPICS-compatible interfaces. This layer ensures scalable, reliable, and synchronized transport of data between experimental stations and computational resources.

The Data Management and Storage layer forms the digital backbone of the architecture. Real-time caches support low-latency operations, while object-based data lakes provide scalable storage for raw experimental outputs. Processed data repositories, relational databases, and metadata catalogs facilitate efficient retrieval, provenance tracking, and long-term archival.

 % \pagebreak
 % \newpage

Building upon this foundation, the Data Processing and Analytics layer performs both real-time and offline computations. Stream-processing frameworks support live monitoring and event detection during experiments, whereas batch-processing environments enable large-scale data reduction and retrospective analysis. Preprocessing and feature-engineering operations further convert raw detector signals into structured representations suitable for advanced analytics and machine learning.

The Machine Learning layer constitutes the analytical core of the framework. Supervised, unsupervised, and self-supervised models can be trained to identify hidden patterns, classify experimental states, detect anomalies, estimate physical parameters, and predict experimental outcomes. Predictions generated by trained models are subsequently integrated within the Adaptive Feedback and Control layer, where they can guide experimental optimization, beam-quality stabilization, anomaly mitigation, and automated parameter tuning.

Finally, the User Interfaces and Visualization layer delivers these capabilities to researchers through dashboards, computational notebooks, web portals, automated reporting tools, and alerting systems. Together, the seven layers establish a pathway toward intelligent and increasingly autonomous facility operations.

The key requirement for successful implementation is bidirectional active coupling. Data should not merely be stored; it should be interpreted quickly enough to influence the next measurement, while human operators retain transparent oversight and physical interpretability. For simplicity of presentation we have avoided the multiple directionalities in the data flow in the illustration (Figure~\ref{fig2}) and rather focussed on the basic minimalistic elemental blocks that are necessary to leverage the full potential in an ultrafast facility setting. This architecture is proposed as a facility-level reference model; individual components are already present at different levels of maturity, but their integration varies among facilities and beamlines.

In the following we present the first small steps that have already started to manifest in the context of diverse user campaigns at ELI facilities, that utilise different components of AI-ML for experiments as well as data interpretation.

\section{Representative Machine-Learning Use Cases Across the ELI Ecosystem and Beyond}
\label{sec:eli_case}
\subsection{Example 1: ELI ALPS}

At ELI ALPS, machine learning can support both ultrafast diagnostics and condensed-phase interpretation \cite{Caruso2025}. One example is carrier-envelope-phase estimation from bulk high-harmonic spectra. Instead of relying only on conventional CEP diagnostics, a trained model can exploit the nonlinear mapping between CEP and harmonic spectral structure, enabling compact, non-invasive, in situ CEP tagging for high-repetition-rate environments. A second example concerns ultrafast exciton relaxation and structural reorganization \cite{Kumar2026} in layered Ruddlesden–Popper perovskites \cite{Yin2021}, where experimental observations can be combined with density-functional theory accelerated by machine-learning interatomic potentials. This illustrates a broader role for machine learning in connecting ultrafast measurements with atomistic structural ensembles and excited-state dynamics.

\begin{figure*}
    \centering
    \includegraphics[width=1\linewidth]{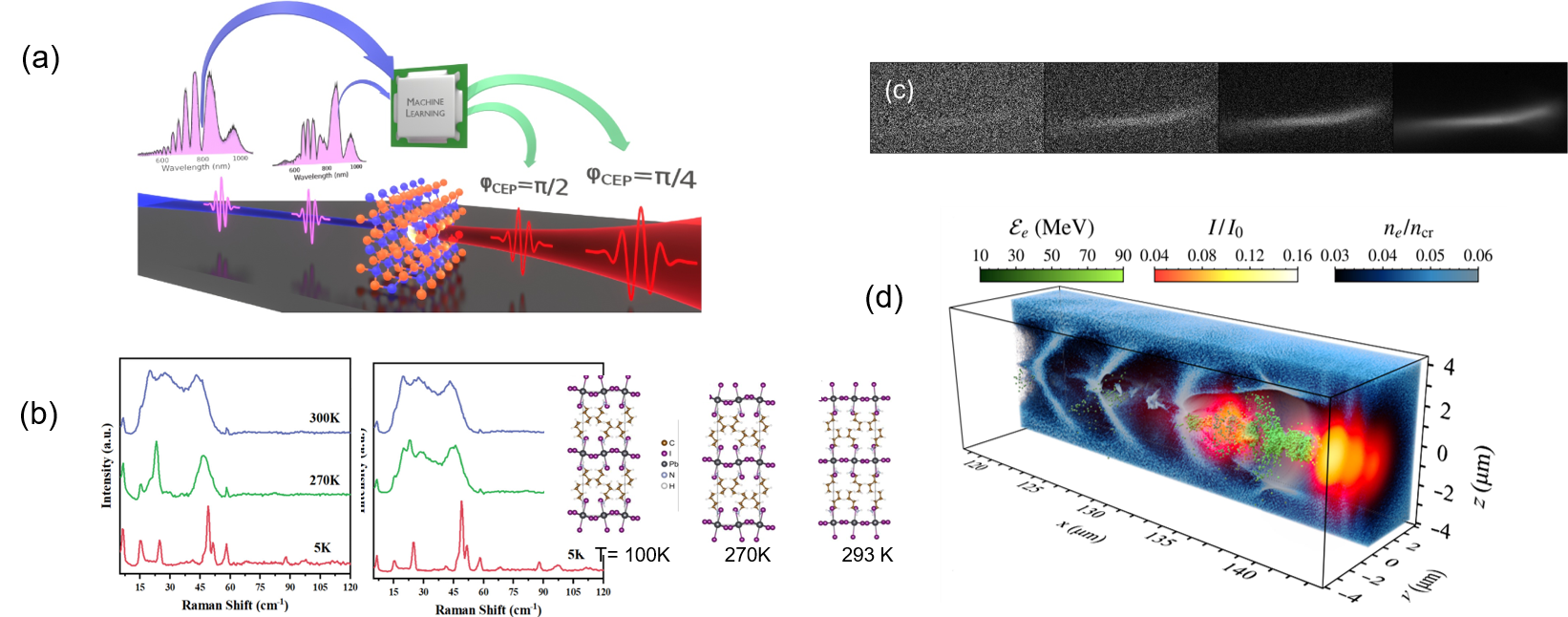}
    \caption{\textbf{Representative examples of machine learning protocols in user and in-house experiments at ELI facilities.} Panels (a) and (b) show studies from ELI ALPS (Szeged, Hungary). (a) Machine-learning-assisted CEP estimation from bulk high-harmonic generation spectra for in situ laser diagnostics. Reproduced with permission from reference \cite{Nagyills2024}. Copyright $2024$ Optica Publishing Group. (b) Accelerated study of ultrafast exciton relaxation and temperature dependent ultrafast structural reorganization (inset) in layered perovskite BA$_2$PbI$_4$ via combined experiments, density functional theory with machine-learning potentials. Panels (c) and (d) present examples from ELI Beamlines (Prague, Czech Republic). (c)  Physics-constrained diffusion-based generation of LWFA electron-spectrometer images from noise conditioned on experimental settings-left to right from noise-dominated input to generated signal. (d) Bayesian optimization with particle-in-cell simulations to identify laser-wakefield acceleration conditions maximizing electron beam energy. Adapted from reference \cite{Valenta2025}. Copyright $2025$ American Physical Society under \href{https://creativecommons.org/licenses/by/4.0/}{CC Licence}. Together, these examples highlight the diverse roles of machine learning in ultrafast science, from diagnostics and materials modeling to experiment optimization and data analysis.}
    \label{fig3}
\end{figure*}

Panels Figure~\ref{fig3}(a) and (b) illustrate two representative examples from ELI ALPS that highlight the growing role of machine learning across distinct stages of the ultrafast science workflow. In panel (a), machine learning is employed for CEP estimation of few-cycle mid-infrared laser pulses using high-harmonic spectra generated in bulk materials \cite{Nagyills2024}. Rather than relying exclusively on conventional CEP diagnostics such as f–2f interferometers or stereo-above-threshold ionization measurements, the approach exploits the complex relationship between CEP and the spectral characteristics of high-harmonic generation. By training a machine learning model on experimentally measured and simulated harmonic spectra, CEP information can be inferred rapidly and non-invasively, enabling compact in situ CEP tagging and providing a practical solution for high-repetition-rate experimental environments. In this investigation first, the feasibility of such tagging is tested on simulated harmonic spectra from a thin ZnO crystal, where the supervised regression and ensemble based ML models are optimised. Then relying on the estimations on the number of necessary training data it is experimentally demonstrated that solid HHG spectra are a good indicator of the relative CEP of the driving laser. Such techniques can be further extended to facilitate challenging experiments within the few-cycle-driven relativistic domain, opening the possibility of correlating spectral signatures with specific pulse control parameters \cite{DeMarco2023,Shirozhan2026,Lamprou2021,Mondal2022}.

Panel Figure~\ref{fig3}(b) demonstrates a complementary application of machine learning in the study of ultrafast exciton relaxation dynamics and structural reorganization in the layered Ruddlesden–Popper perovskite BA$_2$PbI$_4$ \cite{Kumar2026}. Here, experimental observations are combined with density functional theory calculations accelerated through machine-learning-based interatomic potentials, allowing efficient exploration of temperature-dependent structural configurations and their influence on excited-state relaxation pathways. Together, these examples illustrate the breadth of machine learning applications in facility settings, ranging from real-time laser diagnostics and experimental control to the acceleration of computational materials modeling and interpretation of complex ultrafast phenomena.
%%%%%%%%%%%%%%%%%%%%%%%%%%%%%%%%%%%%%%%%%%%%%%%%%%%%%%%%%%%%%%%%%%%%%%%%%%%%%%%%%%%%%%%%%%%%%%%%%%%%%%%%%%%%%%%
%%%%%%%%%%%%%%%%%%%%%%%%%%%%%%%%%%%%%%%%%%%%%%%%%%%%%%%%%%%%%%%%%%%%%%%%%%%%%%%%%%%%%%%%%%%%%%%%%%%%%%%%%%%%%%%
\subsection{Example 2: ELI Beamlines}

At ELI Beamlines, machine learning naturally enters laser tuning, pulse shaping, experiment planning, diagnostics, and data analysis. Bayesian optimization and reinforcement learning can reduce the number of shots required to optimize temporal pulse profiles or secondary-source yield. In laser-wakefield acceleration, Bayesian optimization coupled to particle-in-cell simulations can explore high-dimensional laser–plasma parameter spaces more efficiently than brute-force scans. Physics-constrained diffusion models can generate electron-spectrometer images and corresponding electron-energy spectra conditioned on experimental parameters.

Automated feedback framework was built on the L1 Allegra pump chain, to adjust dispersion terms and optimize the temporal pulse profile in real time. Two algorithms were compared. A Nelder–Mead (NM) simplex algorithm gave rapid local convergence but was prone to getting stuck in local minima. Bayesian optimization (BO) significantly improved global exploration and shortened convergence, especially when compensating for higher-order dispersion terms. BO converged in fewer than 30 iterations and produced a closer match to the target Gaussian pulse shape than NM when, in addition to group delay, third order, and fourth order dispersion, the fifth-order spectral-phase term was included as a control parameter \cite{Matuka2025}. Two reinforcement-learning (RL) approaches at ELI Beamlines tackle the same problem from a sequential-control point of view. The first approach, TempoRL trains a deep RL policy to act directly on the stretcher’s spectral-phase coefficients and reports near-transform-limited pulses \cite{TempoRL}. The second approach, Shaping Laser Pulses with RL learns a policy from Frequency Resolved Optical gating (FROG) data (no intermediate reconstruction), uses bounded one-step actions for machine safety, and adopts domain randomization for robustness across changing conditions \cite{RL_Davorin}. Both RL approaches have so far been demonstrated only in simulation and are intended for future implementation on the L1 laser system.

Machine learning has also been applied in real experimental campaigns. At the L3-HAPLS ELIMAIA beamline, a closed-loop framework combining Bayesian optimization with neural-network surrogate models was deployed  \cite{BL-News}   in high-repetition-rate ion-acceleration experiments to maximize ion yield via adaptive pulse shaping. Operating in real time, it continuously updated control parameters from experimental feedback, demonstrating practical AI-assisted optimization under user-facility constraints where beam time is limited and efficient exploration of parameter space is essential.

Beyond facility control, ML is increasingly integrated with simulations and diagnostics. Bayesian optimization coupled with particle-in-cell (PIC) simulations has been used to identify optimal conditions for laser-wakefield acceleration, enabling efficient exploration of multidimensional laser–plasma parameter spaces while reducing computational cost \cite{Valenta2025}. 

In parallel, researchers from the ALFA LWFA team at ELI Beamlines adapted the Denoising Diffusion Implicit Model (DDIM) framework \cite{DDIM2020} to develop a physically constrained model (PCDDIM) for conditional generation of electron-spectrometer images \cite{Jech2025}. Starting from pure noise, the model iteratively generates a realistic spectrometer image conditioned on experimental settings. The denoiser is conditioned on experimental settings (e.g., laser energy, gas pressure) and trained with simple physics rules to keep the integrated spectrum consistent as noise is added/removed, and to forbid unphysical signal. Compared with a baseline DDIM, this PIDDIM version produces spectra whose probability distribution over energy more closely matches real data. The model also produced physically plausible extrapolations beyond the training parameter range, although further validation using PIC simulations is required. Validated in the ALFA L1-Allegra kHz LWFA program (up to $50~M~eV$), it is well-suited for pre-experimental design (e.g., seeding BO/RL with
realistic priors) to widen the search space while saving beam time  \cite{Jech2025}. Panels Figure~\ref{fig3}(c) and (d) illustrate these two directions: (c) demonstrates diffusion-based conditional generation of LWFA electron-spectrometer images conditioned on experimental settings, while (d) shows BO-guided PIC-based optimization of electron cut-off energy.

\subsection{Example 3: ELI-NP}

At ELI-NP, beam-profile classification and anomaly detection provide practical examples of facility-level decision support. Supervised and transfer-learning models can classify near-field, far-field, multi-spot, and noise images with high accuracy, while feature-based anomaly detection can identify beam instabilities or short-lived deviations in front-end laser diagnostics. Such applications are not merely convenient; they represent the foundation for predictive maintenance, operator assistance, and reproducible machine-state characterization.

The beam-profile classification study \cite{Gaciu2024} constructed a labeled image dataset comprising 14,606 near-field (NF), far-field (FF), multi-spot (``wedge''), and noise frames acquired from the ELI-NP petawatt (PW) laser system and benchmarked several machine-learning classifiers. A simple artificial neural network (ANN) enabled rapid NF-FF classification, while an unsupervised autoencoder clustered NF, FF, and wedge beam profiles with an accuracy of approximately $85\%$. Transfer-learning approaches based on ResNet-18 and GoogLeNet for four-class classification achieved accuracies of approximately $99.8$-$99.9\%$ on held-out test datasets. The authors proposed deploying these models as real-time decision-support tools for facility operators.

In this study \cite{Imran2025}, the front-end laser beam was monitored by extracting feature time series, including centroid drift, intensity mean and variance, Sobel edge sharpness, and ellipticity/roundness metrics, followed by anomaly detection using an Isolation Forest algorithm. Analysis of a 100-image sequence acquired over approximately 4 hours showed stable beam pointing, with centroid fluctuations limited to only a few pixels, while the model consistently identified short-lived anomalous events (e.g., around image 10) across multiple feature channels. Since inference required only milliseconds per frame, the approach demonstrated the feasibility of real-time integration for quality assurance and predictive maintenance applications.
%%%%%%%%%%%%%%%%%%%%%%%%%%%%%%%%%%%%%%%%%%%%%%%%%%%%%%%%%%%%%%%%%%%%%%%%%
\begin{table}[tbh!]
% \small
\large
%\newcolumntype{L}{>{\varwidth[c]{\linewidth}}l<{\endvarwidth}}
%\newcolumntype{M}{>{$}l<{$}}
\caption{Comparison of ELI machine-learning applications discussed in Section~\ref{sec:eli_case}}
\newcolumntype{C}{>{\centering\arraybackslash}c}
\centering
\begin{adjustbox}{max width=1\columnwidth}
%\resizebox{0.95\textwidth}{!}{%
% \begin{tabular}{|p{0.15 \linewidth}|p{0.28 \linewidth}|p{0.32 \linewidth}|p{0.2 \linewidth}|p{0.22 \linewidth}|}
\begin{tabular}{ p{0.24 \linewidth} p{0.36 \linewidth} p{0.3 \linewidth} p{0.26 \linewidth} p{0.36 \linewidth} }
\hline
\rowcolor{white} \textbf{Facility/application} & \textbf{Dataset} & \textbf{ML method} & \textbf{Primary output} & \textbf{Experimental maturity}  \\ 
\hline

\rowcolor{white} ELI ALPS: CEP estimation from bulk HHG spectra & Experimentally measured (100 per CEP) and simulated bulk-HHG spectra ($\sim800$) & Supervised regression and ensemble ML model-based CEP inference & Shot-resolved CEP estimate from the harmonic spectrum & Experimentally demonstrated diagnostic; proposed for compact in situ tagging \\
\hline

\rowcolor{white} ELI ALPS: $BA_{2}PbI_{4}$ structural reorganization and exciton relaxation & Temperature-dependent experiments plus phase-specific DFT structures at 100, 270, and 293 K; no ML is used for the experimental part & DFT-anchored computational workflow; ML interatomic potential or surrogate used to accelerate structural sampling & Temperature-dependent structural ensembles and interpretation of relaxation pathways & Experimental-computational study; ML component is an offline computational accelerator \\
\hline

\rowcolor{white} ELI Beamlines: temporal pulse-shape optimization & FROG-reconstructed temporal profiles acquired during closed-loop optimization; BO converged in fewer than 30 experimental evaluations in the reported run & Nelder-Mead and Bayesian optimization & Optimized compressor dispersion and temporal pulse profile & Experimentally demonstrated closed-loop optimization on the L1 Allegra pump chain \\
\hline

\rowcolor{white} ELI Beamlines: reinforcement-learning based pulse shaping & Simulated pulse-control environments and simulated temporal-profile/FROG feedback; no experimental dataset & Deep reinforcement learning, including TempoRL and image-to-action policies & Control policy for spectral-phase coefficients or pulse-shaping actions & Simulation demonstration; prospective facility implementation \\
\hline

\rowcolor{white} ELI Beamlines: ion-yield optimization at L3-HAPLS ELIMAIA & High-repetition-rate experimental feedback from an ion-acceleration campaign; More than 4000 experimental shots acquired during the L3-HAPLS/ELIMAIA campaign & Bayesian optimization with neural-network surrogate models & Pulse-shaping settings maximizing ion yield & Experimentally demonstrated closed-loop campaign comprising more than 4000 shots \\
\hline

\rowcolor{white} ELI Beamlines: LWFA optimization & 3D Particle-in-cell simulation outputs generated iteratively during BO for self-guided and channel-guided LWFA; no experimental data & Bayesian optimization coupled to PIC simulations & Laser and plasma parameters predicted to maximize electron cut-off energy & Theoretical and numerical study; not experimentally demonstrated \\
\hline

\rowcolor{white} ELI Beamlines: Generative modeling of LWFA electron-spectrometer images & Experimentally measured electron-spectrometer images from 22 LWFA experiments, spanning laser energies of $12-26$ mJ, gas-jet pressures of $10-38$ bar, and acquisition times of $10-40$ ms & Physically constrained denoising diffusion implicit model (PCDDIM) & Generated electron-spectrometer images and corresponding electron-energy spectra & Offline generative-model study trained and cross-validated on experimental ALFA data; intended for simulation and experiment planning, not deployed as a diagnostic or facility-control system \\
\hline

\rowcolor{white} ELI-NP: beam-profile classification & 14,606 near-field, far-field, multi-spot, and noise images & ANN, unsupervised autoencoder, ResNet-18, and GoogLeNet transfer learning & Beam-image class & Experimentally demonstrated offline classification; proposed real-time operator support \\
\hline

\rowcolor{white} ELI-NP: front-end beam anomaly detection & 100 images acquired over approximately 4 h & Feature extraction plus Isolation Forest & Anomaly score and identification of abnormal beam states & Experimental proof of feasibility; millisecond-scale inference reported; decision support rather than autonomous maintenance \\
\hline

\end{tabular}%}	
\end{adjustbox}
\vskip -0.25cm	
\label{table:eli_ml}	
\end{table}
%%%%%%%%%%%%%%%%%%%%%%%%%%%%%%%%%%%%%%%%%%%%%%%%%%%%%%%%%%%%%%%%%%%%%%%%%%%%%%
The Table~\ref{table:eli_ml} distinguishes the application, dataset, machine-learning method, output, and experimental maturity of the representative ELI examples discussed in Section~\ref{sec:eli_case}. Dataset sizes are reported only where the cited study or manuscript provides a quantitative value. In addition, in Table~\ref{table:eli_ml} we have clearly distinguished between experimentally demonstrated applications, and simulation-based studies.

\subsection{Broader International Landscape}
\label{international}
The ELI examples form part of a wider transition from retrospective analysis to shot-resolved inference, adaptive acquisition, and bounded control at ultrafast user facilities. At X-ray free-electron lasers, virtual diagnostics infer photon properties from accelerator observables: at LCLS, machine learning predicted photon energy near $530~eV$ with mean errors below $0.3~eV$, two-pulse delay below $1.6~fs$, and spectral shape with $97\%$ mean agreement; a later LCLS study used approximately 10,000 shots and 192 electron-beam and undulator parameters to predict single-shot spectra and wavefronts \cite{SanchezGonzalez2017,Li2023}. At European XFEL, an uncertainty-aware virtual spectrometer improved non-invasive spectral resolution by up to approximately $40\%$ and was integrated for quasi-real-time and offline use \cite{FerreiradeLima2024}. These results closely parallel the ELI ALPS CEP and reconstruction cases: learned diagnostics supplement slower, invasive, or incomplete measurements, but must remain anchored to traceable calibration and distribution-shift monitoring.

Equally relevant international developments occur in ultrafast condensed-matter and materials science. Gutberlet et al. used a three-layer feed-forward network to exploit correlations in high-harmonic spectra and achieved up to tenfold better noise suppression in XUV transient absorption than conventional referencing, enabling weaker electronic and lattice signals to be resolved \cite{Gutberlet2023}. Chen et al. combined Au/Si ultrafast electron diffraction with scientific machine learning to invert on the order of $10^{3}$ coupled Boltzmann transport equations and recover frequency-dependent interface transmission, relaxation times, and real-space, real-time phonon transport \cite{Chen2022}. At SLAC MeV-UED, multi-objective Bayesian active learning mapped Pareto trade-offs among electron-probe properties with substantially fewer measurements than a grid scan, providing an experimentally demonstrated route to adaptive optimization of instruments used for ultrafast structural dynamics \cite{Ji2024}. In coherent X-ray scattering, the AI-NERD framework applied unsupervised deep learning to experimental XPCS data to classify heterogeneous nonequilibrium relaxation without predefined physical labels and to connect microscopic dynamics with macroscopic properties \cite{Horwath2024}. These studies broaden the ELI comparison from source control to noise suppression, inverse physics, latent-state discovery, and information-efficient experiment design.

For nonlinear laser-plasma experiments, complementary demonstrations at Gemini, DESY, and BELLA establish the maturity of closed-loop optimization and predictive stabilization. Bayesian optimization at Gemini varied up to six laser and plasma controls, with a four-dimensional search converging in approximately 200 shots and a six-dimensional search increasing betatron X-ray yield fivefold \cite{Shalloo2020}. At DESY, PIC-informed Bayesian optimization produced sub-percent energy spread at 254 MeV and identified a regime yielding below $5\%$ rms energy spread for $90\%$ of shots \cite{Jalas2021}. At BELLA, predictive control based on a high-rate pilot beam reduced pointing jitter by $65\%$ and $47\%$ in the two transverse directions at $1~Hz$ \cite{Amodio2025}. Across these materials and plasma examples, the common challenges are nonlinear response, limited beam time, heterogeneous data, drift, damage or safety constraints, and incomplete diagnostics. ELI is distinctive because a distributed infrastructure must support attosecond spectroscopy, ML-accelerated materials interpretation, petawatt laser-plasma experiments, secondary sources, and nuclear photonics. The transferable lesson is therefore not a universal model, but a common deployment strategy based on synchronized metadata, uncertainty-aware inference, cross-campaign validation, and staged progression from offline analysis to advisory and bounded closed-loop operation.

Table~\ref{table:international_ml} compares these studies by task, quantitative evidence, maturity, and relevance to ELI. Collectively, they show that ML in ultrafast facilities is progressing along complementary paths, from virtual diagnostics and materials-dynamics inference to adaptive optimization and bounded experimental control. These differences in task and maturity require correspondingly different standards of uncertainty, interpretability, transferability, cybersecurity, and operational authorization.

%%%%%%%%%%%%%%%%%%%%%%%%%%%%%%%%%%%%%%%%%%%%%%%%%%%%%%%%%
\begin{table}[t!]
% \small
\large
%\newcolumntype{L}{>{\varwidth[c]{\linewidth}}l<{\endvarwidth}}
%\newcolumntype{M}{>{$}l<{$}}
\caption{International ML developments relevant to the ELI scientific use cases}
\newcolumntype{C}{>{\centering\arraybackslash}c}
\centering
\begin{adjustbox}{max width=1\columnwidth}
%\resizebox{0.95\textwidth}{!}{%
% \begin{tabular}{|p{0.15 \linewidth}|p{0.28 \linewidth}|p{0.32 \linewidth}|p{0.2 \linewidth}|p{0.22 \linewidth}|}
\begin{tabular}{ p{0.2 \linewidth} p{0.28 \linewidth} p{0.36 \linewidth} p{0.26 \linewidth} p{0.3 \linewidth} }
\hline
\rowcolor{white} \textbf{Facility/platform} & \textbf{Application/ML method} & \textbf{Quantitative evidence} & \textbf{Maturity} & \textbf{Relevance to ELI}  \\ 
\hline

\rowcolor{white} LCLS, USA & Virtual diagnostics for XFEL pulse
properties & $<0.3$ eV photon-energy error; $<1.6$ fs delay error; $97\%$ spectral agreement \cite{SanchezGonzalez2017} & Experimental, shot-resolved & CEP tagging and virtual diagnostic
calibration \\
\hline

\rowcolor{white} European XFEL, Germany & Uncertainty-aware virtual spectrometer & Up to $~40\%$ spectral-resolution improvement \cite{Li2023} & Operationally deployed & Uncertainty, drift monitoring, and user-facing inference \\
\hline

\rowcolor{white} HHG-XUV transient absorption, Germany & Feed-forward neural-network denoising & Up to tenfold noise suppression \cite{Gutberlet2023} & Experimental spectroscopy & Weak-signal electron and lattice dynamics \\
\hline

\rowcolor{white} Au/Si MeV-UED, USA & Scientific-ML inversion of phonon
transport & Inversion of $O(10^{3})$ coupled transport equations \cite{Chen2022} & Experimental data plus physicsconstrained inversion & Latent microscopic parameters from ultrafast measurements \\
\hline

\rowcolor{white} SLAC MeV-UED, USA & Multi-objective Bayesian active learning & Pareto mapping with substantially fewer measurements than grid scanning \cite{Ji2024} & Experimental online tuning & Joint optimization of resolution, signal, stability, and damage \\
\hline

\rowcolor{white}APS XPCS / AI-NERD, USA & Unsupervised deep learning of relaxation dynamics & Automated classification without prior physical labels \cite{Horwath2024} & Experimental materials analysis & Discovery of heterogeneous transient regimes \\
\hline

\rowcolor{white} Gemini, UK & Bayesian optimization of LWFA & ~200 shots for 4D convergence;
fivefold X-ray-yield increase in 6D \cite{Shalloo2020} & Experimental closed loop & Shot-efficient nonlinear laser-plasma optimization \\
\hline

\rowcolor{white} DESY, Germany & PIC-informed Bayesian optimization & Sub-percent spread at 254 MeV; $<5\%$ rms for $90\%$ of shots \cite{Jalas2021} & Simulation-informed experiment & Simulation-to-experiment transfer and robustness\\
\hline

\rowcolor{white} BELLA, USA & Neural predictive laser-pointing control & Jitter reduced by $65\%$ and $47\%$ at 1 Hz \cite{Amodio2025} & Experimental predictive control & High repetition rate proxy diagnostics for low repetition rate high power lasers \\
\hline

\end{tabular}%}	
\end{adjustbox}
\vskip -0.25cm	
\label{table:international_ml}	
\end{table}
%%%%%%%%%%%%%%%%%%%%%%%%%%%%%%%%%%%%%%%%%%%%%%%%%%%%%%%%%%%%
\section{Trustworthy and Transferable Deployment During User Campaigns}

Trustworthy deployment therefore requires more than high predictive accuracy on a held-out dataset. While these studies demonstrate the feasibility of accurate and rapid ML inference for specific diagnostics and operational tasks, moving from such individual demonstrations to reliable deployment across live user campaigns introduces additional requirements concerning uncertainty quantification, model interpretability, transferability, and data standardization. It is important that ML predictions used for operational decision-making are accompanied by appropriate uncertainty estimates, validated on independent runs and under changed operating conditions, with out-of-distribution cases triggering warnings or fallback to conventional diagnostics \cite{FerreiradeLima2024}. Model interpretability should further help establish whether predictions are driven by physically meaningful variables rather than experimental artefacts. This can be assessed using feature attribution, sensitivity or counterfactual analysis, together with comparisons against established physical relationships, simulations, or independent diagnostics \cite{Grojean2022,Karniadakis2021}. Transferability should be demonstrated across different runs, campaigns, beamlines, detector configurations, and operating regimes, while standardized data and metadata (including consistent units, calibration information, preprocessing, software/model versions, and data-quality indicators) are essential for reproducibility and reuse following FAIR principles \cite{Wilkinson2016}. In the heterogeneous ELI environment, these requirements suggest that transferable components are more likely to be data schemas, interfaces, representations, and validation protocols than unchanged end-to-end models.

These requirements also naturally increase the complexity and connectivity of the operational technology infrastructure. Supporting standardized data flows, uncertainty monitoring, model validation, and interaction with machine controls may require additional interfaces. These include gateways, data pathways, and software services, which increase the potential attack surface and make cybersecurity an essential part of ML deployment. Appropriate measures include network segmentation, authenticated and least-privilege access, protected communications, software and model version control, audit logging, and independent safety interlocks \cite{Stouffer2023,CF2024}. A staged deployment pathway, progressing from retrospective validation and shadow-mode testing to human-approved advisory operation and, where justified, bounded closed-loop control, can provide a practical approach for introducing ML into user campaigns while maintaining scientific, operational, and cybersecurity safeguards.

These safeguards do not constitute barriers to adaptive experimentation; they define the conditions under which it becomes scientifically credible. By matching model authority to demonstrated maturity, ELI can introduce ML incrementally, beginning with offline interpretation and operator support and progressing toward bounded feedback only when uncertainty, transferability, latency, and fail-safe behavior have been validated. This graded approach provides the foundation for the conclusions drawn below.

\section{Conclusions}

Machine learning is emerging as an enabling methodology for facility-scale ultrafast science, where high-dimensional control spaces, heterogeneous diagnostics, shot-to-shot variability, and limited beam time increasingly constrain conventional experimental workflows. The examples discussed across the ELI ecosystem demonstrate complementary roles for AI-ML, including carrier-envelope-phase inference, accelerated modeling of structural and excitonic dynamics, physics-constrained generative modeling of diagnostic data, beam-state monitoring, and efficient optimization of high-power laser and laser-plasma experiments. Importantly, these applications span different levels of maturity, from simulation-based studies and offline interpretation to experimentally demonstrated diagnostics and bounded closed-loop optimization.

The broader international landscape confirms that these challenges are not unique to ELI. Developments at free-electron lasers, ultrafast electron-diffraction facilities, and time-resolved materials-science platforms show that ML can enhance measurement sensitivity, recover latent physical information, identify nonequilibrium dynamical regimes, and reduce the number of measurements required for optimization. ELI’s distinctive opportunity lies in integrating such capabilities across attosecond science, ultrafast condensed matter, laser-plasma physics, secondary-source generation, and nuclear photonics within a distributed research infrastructure.

Achieving this integration requires more than predictive accuracy. Models must be connected to calibrated measurements, supported by synchronized metadata and traceable provenance, validated across changing experimental conditions, and accompanied by uncertainty estimates and physically meaningful checks. Their operational authority should increase only through staged validation, from offline analysis and operator support to bounded feedback with independent safeguards. Within this framework, ML does not replace physical understanding or scientific judgment. It provides the speed, scalability, and adaptive capability needed to convert increasingly complex experimental data into reproducible insight and, ultimately, to enable trustworthy data-driven discovery in ultrafast science.  Looking forward, facilities such as the Extreme Light Infrastructure are poised to evolve into AI-augmented, closed-loop research ecosystems, where intelligent control, real-time inference, and adaptive experimentation will jointly enable discovery beyond the reach of conventional human-guided operation. In the longer term, integrating these capabilities with robust facility control could enable progressively autonomous ultrafast experiments that adapt their measurement strategies in real time while remaining bounded by quantified uncertainty, physical constraints, operational safeguards, and transparent human oversight.

\section*{Acknowledgements}

This work was partially supported by the 4U-PI project funded by the European Union's Horizon Europe Research and Innovation Programme under Grant Agreement No. 101290656. Views and opinions expressed are however those of the author(s) only and do not necessarily reflect those of the European Union or the Agency. Neither the European Union nor the granting authority can be held responsible for them. The ELI ALPS project (GINOP-2.3.6-15-2015-00001) is supported by the European Union and co-financed by the European Regional Development Fund.

% \printbibliography
\end{document}